\documentclass[ ,final 
  ] {aipproc}

\layoutstyle{8x11single}

\begin{document}

\title{Digging for the Truth: Photon Archeology with GLAST}

\classification{98.35Ac, 98.70Rz, 98.70Vc}
                \keywords {gamma-rays, absorption, star formation, 
                          background radiation}

\author{F.  W. Stecker}{  address={NASA Goddard  Space  Flight Center,
  Greenbelt, MD 20771, USA} }



\begin{abstract}

Stecker, Malkan  and Scully,  have shown how  ongoing deep  surveys of
galaxy  luminosity   functions,  spectral  energy   distributions  and
backwards  evolution models  of star  formation rates  can be  used to
calculate  the  past history  of  intergalactic  photon densities  for
energies from 0.03 eV to the  Lyman limit at 13.6 eV and for redshifts
out to 6 (called here the intergalactic background light or IBL). From
these calculations of  the IBL at various redshifts,  they predict the
present  and  past  optical  depth  of the  universe  to  high  energy
$\gamma$-rays owing  to interactions with  photons of the IBL  and the
2.7 K  CMB. We  discuss here  how this proceedure  can be  reversed by
looking for sharp cutoffs in the spectra of extragalactic $\gamma$-ray
sources  such as  blazars at  high redshifts  in the  multi-GeV energy
range with {\it GLAST} (Gamma-Ray Large Are Space Telescope).  
By determining  the cutoff energies of sources with
known redshifts,  we can  refine our determination  of the  IBL photon
densities in the past, {\it i.e.}, the {\it archeo-IBL}, and therefore
get a better  measure of the past history of  the {\it total} star formation
rate.  Conversely,  observations of sharp  high energy cutoffs  in the
$\gamma$-ray  spectra of  sources  at unknown  redshifts  can be  used
instead of spectral lines to give a measure of their redshifts.



\end{abstract}

\maketitle


\section{Introduction}

The potential importance of the photon-photon pair production process,
$\gamma \gamma  \rightarrow e^+ e^-$, in high  energy astrophysics has
been realized  for over 40 years  \cite{ni}.  It was  pointed out that
owing to interactions with the 2.7 K CMB, the universe would be opaque
to $\gamma$-rays  of energy above  100 TeV at  extragalactic distances
\cite{gs, j}.  If one  considers cosmological and redshift effects, it
was  further  shown that  photons  from  a  $\gamma$-ray source  at  a
redshift $z_{s}$  would be  significantly absorbed by  pair production
interactions with the CMB above an energy $\sim 100(1+z_{s})^{-2}$ TeV
\cite{st69, fs}.

Following  the discovery  by  the  {\it EGRET}  team  of the  strongly
flaring  $\gamma$-ray  blazar   3C279  at  redshift  0.54  \cite{h92},
Stecker, de Jager and Salamon \cite{sds} proposed that one can use the
predicted pair production absorption  features in blazars to determine
the intensity  of the infrared portion  of the IBL,  provided that the
intrinsic spectra  of blazars  extends to TeV  energies. It  was later
shown that the  IBL produced by stars in galaxies  at redshifts out to
$\sim 2$ would make the universe  opaque to photons above an energy of
$\sim 30$  GeV emitted  by sources  at a redshift  of $\sim  2$, again
owing to pair production interactions \cite{mp, ss98}.

In Ref.\cite{sms} this approach was expanded by using recent data from
the {\it  Spitzer} infrared observatory  the {\it Hubble}  deep survey
and {\it GALEX}  to determine the photon density of  the IBL from 0.03
eV  to  the Lyman  limit  at  13.6 eV  for  redshifts  out  to 6  (the
``archeo-IBL'').\footnote{See also Refs. \cite{tt, k1, k2} and the 
paper of D. H. Hartmann, these proceedings.}   
The results,  giving the  IBL photon  density  as a
function  of  redshift together  with  the opacity  of  the  CMB as  a
function of redshift,  were then used to calculate  the opacity of the
universe to $\gamma$-rays  for energies from 4 GeV to  100 TeV and for
redshifts  from $\sim 0$  to 5.  The results  of this  calculation are
shown  in Figure 1.   They are  given for  two evolution  models, {\it
viz.},  the ``baseline''  (B) model  and the  ``fast  evolution'' (FE)
model. These two  models were chosen to bracket  the plausible history
of the star  formation rates in galaxies which is  the input which has
the largest uncertainty in the calculation, particularly at the higher
redshifts.  The   FE  model  is   favored  by  recent   {\it  Spitzer}
observations \cite{lf,  pg}. It also provides a  better description of
the deep  {\it Spitzer} number counts  at 70 and 160$\mu$m  than the B
model.  However, {\it  GALEX} (Galaxy Evolution Explorer) observations
indicate  that the  redshift evolution  of  UV radiation  may be  more
consistent  with  the  B  model  \cite{sch}. The  {\it  Spitzer  IRAC}
(Infrared Array Camera) counts can  be best fit with an evolution rate
between  these two  models.   One way  of  understanding the  somewhat
smaller redshift evolution  of the star formation rate  implied by the
{\it  GALEX} UV  observations {\it  vs.} that  obtained from  the {\it
Spitzer}  IR  observations  is  that  the effect  of  dust  extinction
followed by IR reradiation increases with redshift \cite{bu}.

It has been argued that TeV $\gamma$-ray observations of the blazars
1ES1101-232 and H 2356-309 place an upper limit on the present IBL
of $14\pm4$ nW m$^{-2}$sr$^{-1}$ in the wavelength range 1--2 $\mu$m
\cite{ah06}, based on the assumption that the intrinsic photon spectra
of these sources in the energy range of observation cannot have a 
spectral index which is flatter than 1.5. Should this assumption be 
true, it would tend to favor model B \cite{ss06}.  

The results of Ref. \cite{sms} predict that the universe will 
become opaque to
$\gamma$-rays for  sources at the  higher redshifts at  somewhat lower
$\gamma$-ray energies  than those given in Ref.   \cite{ss98}. This is
because the  newer deep surveys  have shown that there  is significant
star formation  out to redshifts  $z \ge 6$ \cite{bu04,  bo}), greater
than  the value of  $z_{max} =  4$ assumed  in Ref.  \cite{ss98}. This
conclusion is also supported by  recent {\it Swift} observations of the
redshift distribution of GRBs \cite{tj}.

\begin{figure}
  \includegraphics[height=.3\textheight]{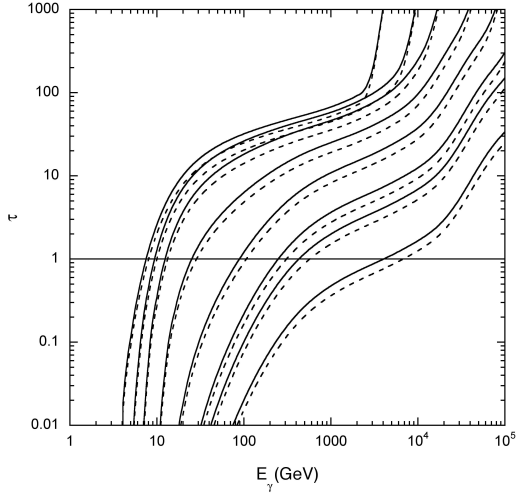}
  \caption{Optical depth of the universe to $\gamma$-rays from interactions
with photons from the EBL and 2.7K CMB for various redshifts, $z$. The solid 
lines are for the FE model
and the dashed lines are for the B model. The curves shown are for (from top
to bottom) $z$ = 5.0, 3.0, 2.0, 1.0, 0.5, 0.2, 0.117, 0.03\protect\cite{sms}.}
\end{figure}

Stecker, Malkan and Scully \cite{sms} found that  the function 
$\tau (E_{\gamma}, z)$  shown in Fig. 1 can be very well approximated
by the analytic form

\begin{equation}
\log \tau = Ax^4+Bx^3+Cx^2+Dx+E 
\end{equation}

\noindent over  the range  $0.01 <  \tau < 100$  where $x  \equiv \log
E_{\gamma}$ (eV).  The  coefficients A through E are  given in Table 1
for  various redshifts \cite{cor}. (This is the corrected form of the
original table given in Ref. \cite{sms}.) 

\begin{table}
\begin{tabular}{lrrrrr}
\hline

$z$ & $A$ & $B$ & $C$ &  $D$ & $E$\\
0.03 & -0.020228 & 1.28458 & -29.1498 & 285.131 & -1024.64 \\ 
~&  -0.020753  &    1.31035   &  -29.6157   &   288.807  &   -1035.21 \\
0.117 & 0.010677 & -0.238895 & -1.004 & 54.1465 & -313.486 \\ 
~& 0.022352  &   -0.796354 &    8.95845 &    -24.8304  &   -79.0409 \\
0.2 & 0.0251369 & -0.932664 & 11.4876 & -45.9286 & -12.1116 \\
~&  0.0258699  &  -0.960562 &    11.8614  &   -47.9214   &  -8.90869 \\
0.5 & -0.0221285 & 1.31079 & -28.2156 & 264.368 & -914.546 \\ 
~&  0.0241367  &  -0.912879 &    11.7893 &    -54.9018   &   39.2521 \\
1.0 & -0.175348 & 8.42014 & -151.421 & 1209.13 & -3617.51 \\ 
~& -0.210116  &    10.0006 &    -178.308  &    1412.01  &   -4190.38 \\
2.0 & -0.311617 & 14.5034 & -252.81 & 1956.45 & -5671.36 \\ 
~& -0.397521  &    18.3389  &   -316.916  &    2431.84 &    -6991.04 \\
3.0 & -0.34995 & 16.0968 & -277.315 & 2121.16 & -6077.41 \\
~ & -0.344304 & 15.8698 & -273.942 & 2099.29 & -6025.38\\ 
5.0 & -0.321182 & 14.6436 & -250.109 & 1897.00 & -5390.55 \\
~&-0.28918 &  13.2673 & -227.968 &  1739.11 &  -4969.32 \\

\end{tabular}
\caption{Coefficients for Parametric Fit to $\tau (E_{\gamma},z)$
for various
redshifts for the baseline model (upper row) and fast evolution (lower
row) for each individual redshift. The parametric approximation holds
for $10^{-2} < \tau < 10^2$ and $E_{\gamma} < \sim2$ TeV for all redshifts 
but also up to $\sim$ 10 TeV for redshifts, $z \le 1$ ~\protect\cite{cor}. }
\label{tab:a}
\end{table}

\section{What GLAST Can Do}
 
It can be seen from Figure 1 that for  $\gamma$-ray sources  at the  
higher redshifts
there is a  steeper energy dependence of the optical depth
$\tau  (E_{\gamma})$ near the
energy  where $\tau =  1$.  There  will thus  be a  sharper absorption
cutoff for sources at high redshifts.  It can easily be seen that this
effect is  caused by the  sharp drop in  the UV photon density  at the
Lyman limit.

It is expected that {\it GLAST} will be able to resolve out thousands
of blazars \cite{ss96}.
Because of the strong  energy dependence of absorption in blazar spectra
at the higher redshifts in the multi-GeV range, {\it GLAST} will be able 
to  probe 
the archeo-IBL and thereby probe the early star formation rate.
{\it  GLAST} should  be able  to detect
blazars at known redshifts $z \sim 2$ at multi-GeV energies and determine their critical cutoff energy. A simple
observational technique for probing the archeo-IBL has been proposed in 
Ref. \cite{crr}.
In such ways, 
{\it GLAST} observations  at redshifts $z \ge 2$  and $E_{\gamma} \sim
10$ GeV may complement the  deep galaxy surveys by probing the {\it total}
star formation rate, even that from galaxies too faint to be detected in the
deep surveys. Future {\it GLAST} 
observations in the 5 to 20 GeV energy range may also help to pin down 
the amount of dust extinction in high-redshift galaxies  
by determining the mean density of UV photons 
at the higher redshifts through their absorption effect on the
$\gamma$-ray spectra of high redshift sources.
If the diffuse  $\gamma$-ray background  radiation is  from unresolved
blazars \cite{ss96},  a  hypothesis  which  can  be
independently  tested by {\it  GLAST} \cite{ss99}, absorption will  
steepen the spectrum of this radiation
at $\gamma$-ray energies above $\sim 10$  GeV \cite{ss98}. Thus, {\it GLAST} 
can also aquire  information about the evolution  of the IBL in this way.

Conversely,  observations of sharp  high energy cutoffs  in the
$\gamma$-ray  spectra of  sources  at unknown  redshifts  can be  used
instead of spectral lines to give a measure of their redshifts.



\begin{thebibliography}{99}

\bibitem{ni} A. I. Nikishov, Sov. Phys. JETP {\bf 14}, 373 (1962).

\bibitem{gs} R. J. Gould and G. Schr\'{e}der, Phys. Rev. Letters {\bf 16}, 
252 (1966).

\bibitem{j} J. V. Jelly, Phys. Rev. Letters {\bf 16}, 479 (1966). 

\bibitem{st69} F. W. Stecker, Astrophys. J. {\bf 157}, 507 (1969).

\bibitem{fs} G. G. Fazio and F. W. Stecker, Nature {\bf 226}, 135 (1970).

\bibitem{h92} R. C. Hartman {\it et al.}, Astrophys. J. {\bf 385}, L1 (1992). 

\bibitem{sds} F. W. Stecker, O. C. De Jager and M. H. Salamon, Astrophys. J. 
{\bf 390}, L49 (1992).

\bibitem{mp} P. Madau and E. S. Phinney, Astrophys. J. {\bf 456}, 124 (1996). 

\bibitem{ss98} M. H. Salamon and F. W. Stecker, Astrophys. J. {\bf 493}, 
547 (1998). 

\bibitem{sms} F. W. Stecker, M. A. Malkan and S. T. Scully, Astrophys. J. 
{\bf 648}, 774 (2006).

\bibitem{tt} T. Totani and T. T. Takeuchi, Astrophys. J. {\bf 570}, 470 (2002). 
\bibitem{k1} T. M. Kneiske, K. Mannheim and D. H. Hartmann, Astron. and
Astrophys. 386, 1 (2002). 

\bibitem{k2} T. M. Kneiske {\it et al.}, Astron. and
Astrophys. 413, 807 (2004). 

\bibitem{lf} Le Floc'h {\it et al.}, Astrophys. J. {\bf 632}, 169 (2005). 

\bibitem{pg} P. G. P\'{e}rez-Gonz\'{a}lez {\it et al.} Astrophys. J. 
{\bf 630}, 82 (2005).

\bibitem{sch} D. Schiminovich {\it et al.}, Astrophys. J. {\bf 619}, L47 
(2005). 
\bibitem{bu} D. Burgarella {\it et al.} (GALEX team), in {\it The Dusty 
and Molecular Universe}, ed. A. Wilson (ESA SP-577); Nordwijk: ESA), 141 
(2006).

\bibitem{ah06} F. Aharonian {\it et al.}, Nature {\bf 440}, 1018 (2006).

\bibitem{ss06} F. W. Stecker and S. T. Scully, Astrophys. J. {\bf 652}, L9. 

\bibitem{bu04} A. J. Bunker {\it et al.} Mon. Not. Roy. Ast. Soc. {\bf 355},
374 (2004). 

\bibitem{bo} R. J. Bouwens {\it et al.} Astrophys. J. {\bf 653}, 53 (2006).

\bibitem{tj} N. R. Tanvir and P. Jakobson, Phil Trans Roy. Soc. A, in press,
e-print astro-ph/0701777. 

\bibitem{cor}  F. W. Stecker, M. A. Malkan and S. T. Scully, Astrophys. J. 
{\bf 658}, 1392 (2007).

\bibitem{ss96} F. W. Stecker and M. H. Salamon, Astrophys. J. {\bf 464}, 600. 
(1996). 

\bibitem{crr} A. Chen, L. C. Reyes and S. Ritz, Astrophys. J. {\bf 608}, 686
(2004). 

\bibitem{ss99} F. W. Stecker and M. H. Salamon, {\it Proc. XXVI Intl. Cosmic 
Ray Conf.} ed. D. Kieda, M. Salamon and B. Dingus {\bf 3}, 313 (1999);
e-print astro-ph/9909157.

\end{thebibliography}
\end{document}